# The effect of a radome on submillimeter site testing measurements

Paolo G. Calisse

School of Physics, University of New South Wales, Sydney NSW 2052, Australia
e-mail: pcalisse@phys.unsw.edu.au

**ABSTRACT**

We evaluate the effect that radome transparency has on atmospheric opacity measurements performed by the skydip technique. We show that, except at rather high opacities, it is not sufficient to ignore losses in the radome (or "window") during the data analysis and then subtract them off from the derived atmospheric opacity. Perhaps surprisingly, unless radome transparency is correctly modelled, the atmosphere will appear to have a minimum opacity that is many times greater that the radome losses. Our conclusion is that some previous site studies may have significantly underestimated the quality of the best submillimeter sites, and that the difference between these sites and poorer sites may be much greater that currently believed. We also show that most of the residual 857 GHz opacity at the best sites, currently ascribed to "dry-air opacity", can in fact be just an artefact caused by not properly modelling the radome during the data analysis.

## INTRODUCTION

The skydip technique (Dicke et al. 1946) is used to measure the atmospheric opacity from radio to infrared wavelengths, both for site testing purposes and to calibrate observations. The technique requires a measurement of the sky flux at different elevations, and derives the zenith sky opacity on the assumption that the atmosphere can be modelled as a single slab at uniform temperature.

Skydips have been widely used in the sub-millimeter and millimeter wavelength region (Chamberlin & Bally 1994; Chamberlin, Lane, & Stark 1997; Dragovan et al. 1990; Lane 1998; Matsuo, Sakamoto, & Matsushita 1998; Pardo, Serabyn, & Cernicharo 2001; Radford & Holdaway 1998), because, when observing astronomical sources at these wavelengths, the attenuation introduced by the atmosphere, and its fluctuations, are amongst the main sources of noise, calibration error, and systematic errors.

We have reviewed the different algorithms that have been used for data reduction, which allow for different contributions. In the general case one should include terms related to the sky, such as the cosmic background radiation, or to the telescope and receiver, like spillover, blockage and ohmic losses (Ulich & Haas 1976). With a finite beam size the non-linearity of the relation between the elevation and the airmass number when approaching the horizon should also be considered. This limits the minimum elevation to use for the skydip depending on the instrument beam size (Rohlfs & Wilson 1996) and the geographical characteristics at a particular site.

However, the algorithms traditionally used do not always take into account the effect of absorbing material used for the radome or window[1]. We will show that this effect cannot, in many conditions, be treated as a simple offset to be subtracted later from the sky opacity. Moreover, in particularly good atmospheric conditions the instrument will be unable to obtain a valid and unambiguous value for the sky opacity.

This ambiguity can, however, be fully removed using an appropriate data reduction algorithm, as shown below.

## THE ATMOSPHERIC SINGLE SLAB MODEL

Most authors make use of the standard atmospheric model (in which the atmosphere is considered to be a plane-parallel slab at uniform temperature) to describe the relation between the atmospheric emission and the airmass number $X = 1/\sin(z)$ at elevation $z$. Assuming that one is using a chopping technique, the signal at the detector (expressed in temperature units), $T'(X)$, is:

$$T'(X) = T'_{atm} \cdot (1 - e^{-X \cdot \tau'_0}) - T_{ref} \qquad (1.1)$$

where $\tau'_0$ is the estimated zenith opacity, $T'_{atm}$ the physical temperature of the atmosphere, and $T_{ref}$ the reference temperature against which the chopping is performed.

All the other corrections (cosmic background radiation, spillover, ohmic losses, etc.) have been neglected. A prime will be used to distinguish values obtained neglecting the window transparency from those that properly include this correction.

## OPACITY OBTAINED FROM MEASUREMENTS AT 1 AND 2 AIRMASSES

From eq. (1.1), we obtain at 1 and 2 airmasses:

$$T'(1) = T'_{atm}(1 - e^{-\tau'_0}) - T_{ref} \qquad (2.1)$$

$$T'(2) = T'_{atm}(1 - e^{-2\tau'_0}) - T_{ref} \qquad (2.2)$$

We can analytically solve this system of two equations for $\tau'_0$ and $T'_{atm}$, as follows:

$$y' = \frac{T'(2) - T'(1)}{T'(1) + T_{ref}} \qquad (2.3)$$

$$T'_{atm} = \frac{T'(1) + T_{ref}}{1 - y'} \qquad (2.4)$$

where $y' = e^{-\tau'_0}$ is the atmospheric transmission.

---

[1] In some cases the effect of the radome can be correctly modeled or measured as a component of the antenna beam coupling efficiency.

However, if a radome is used, the detected signal $T(X)$ at a given airmass number $X$ should properly be written as:

$$T(X) = \vartheta_{win}T_{atm}(1-e^{-X\tau_0}) + T_{win}(1-\vartheta_{win}) - T_{ref} \quad (2.5)$$

where $\vartheta_{win}$ is the transparency of the window, and $T_{win}$ its temperature (De Zafra 1995). We have neglected scattering and reflection by the window for the following reasons. Scattering would essentially introduce a spillover term affecting the relation between the model and experimental data, especially at low elevation, where the airmass number changes dramatically for little change of elevation. This deviation is not seen in real data (Calisse et al. 2002). Reflection would introduce a thermal signal in the same way that absorption does, and can be treated simply as another form of absorption as long as all temperatures in the vicinity of the window are comparable. We show below that the window temperature is in fact not critical, and hence this is a good approximation.

For the case of Gore-tex, a fiber widely used as a radome material, these assumptions are also supported by direct measurements.

Within this approximation, we have at 1 and 2 airmasses:

$$T(1) = \vartheta_{win}T_{atm}(1-e^{-\tau_0}) + T_{win}(1-\vartheta_{win}) - T_{ref} \quad (2.6)$$

$$T(2) = \vartheta_{win}T_{atm}(1-e^{-2\tau_0}) + T_{win}(1-\vartheta_{win}) - T_{ref} \quad (2.7)$$

The correct value of the zenith opacity $\tau_0$ can be found substituting the values of $T_{atm}$ obtained from equation (2.6) into eq. (2.7):

$$y = \frac{T(2) - T(1)}{T(1) - T_{win} \cdot (1-\vartheta_{win}) + T_{ref}} \quad (2.8)$$

$$T_{atm} = \frac{T(1) - T_{win}(1-\vartheta_{win}) + T_{ref}}{\vartheta_{win}(1-y)} \quad (2.9)$$

where $y = e^{-\tau_0}$ is now the actual atmospheric transmission.

We will now evaluate the error that is introduced by ignoring radome losses, as in eq. (1.1), substituting the values for $T(1)$ and $T(2)$ obtained from eqs. (2.6) and (2.7) into the approximate solutions (2.3) and (2.4).

We obtain:

$$y' = \frac{1-y}{k-y}y \quad (2.10)$$

$$T'_{atm} = \vartheta_{win}T_{atm}\frac{(k-y)^2}{k-2y+y^2} \quad (2.11)$$

where $k = 1 + \frac{T_{win}}{T_{atm}} \cdot \frac{1-\vartheta_{win}}{\vartheta_{win}}$. The apparent zenith opacity $\tau'_0$ is plotted against the actual opacity $\tau_0$, for typical conditions, in Fig. 1.

In high opacity conditions the difference between the actual and the apparent opacity is independent of the opacity value, as expected. In fact, for $\tau_0 \gg 1$, we can neglect $y$ with respect to 1 in eq. (2.10) and obtain an asymptotic correction for $\tau_0$ and $T_{atm}$:

$$\tau_0 = \tau'_0 - \ln[k] \qquad (2.12)$$

$$T_{atm} = \frac{1}{k\vartheta_{win}} T'_{atm} \qquad (2.13)$$

Note that $k > 1$; that is, neglecting the window opacity always leads to an over-estimate of the atmospheric opacity.

For $\vartheta_{win} = 0.81$, a typical Gore-Tex transmission at 857 GHz, (Radford 2002), $T_{win} = 270K$ and $T_{atm} = 220K$ (typical winter South Pole conditions) we have $\tau_0 = \tau'_0 - 0.25$ and $T_{atm} = 0.96 \cdot T'_{atm}$ for high opacity values.

However, in the more important case of medium and low opacities, the approximate model fails altogether. In fact, there is a minimum value to the apparent opacity that the skydip technique will measure when using a window of transparency $\vartheta_{win}$. This value can be obtained by finding the zeroes of the derivative of $y'$ with respect to $y$:

$$y_{\min} = k - \sqrt{k \cdot (k-1)} \qquad (2.14)$$

For a given radome transparency, as the sky opacity progressively decreases, the approximate model creates first an increased offset, then a reduced sensitivity to opacity changes, and, finally, the apparent opacity actually increases, and the model described by eq. (1.1) generates an ambiguity in the opacity data.

If the correct model is used to solve the system of equations or to fit the data, this problem disappears.

To understand when, in general, the ambiguity can occur, we plotted the relation between the radome transparency and the apparent minimum opacity (see Fig. 2), for two different ratios between the radome temperature $T_{win}$ and the actual atmospheric temperature $T_{atm}$. We find that the results are not very sensitive to the ratio between these two temperatures. This can be explained by the fact that the emission term in eq. (2.5) is usually negligible with respect to the absorption term. So, even in the presence of relatively high (30-50 K) uncertainties in the window temperature we will still obtain valid results.

## OPACITY OBTAINED BY NON-LINEAR FITTING

In the previous section we discuss measurements taken only at 1 and 2 airmasses. However, opacity measurements are often made by observing at many different airmasses and using a non-linear fit to the data. As demonstrated in another paper (Calisse, et al. 2002), we observed an excellent agreement between the parameters obtained with both techniques for large sets of opacity data obtained at Dome C and South Pole.

For data already published in the literature that did not properly account for the radome opacity, we can nevertheless calculate a correction. The correction to be applied will be now slightly sensitive to the details of the sampling (number and values of airmass steps) and so obtaining an analytical expression for the correction will not be possible.

For this reason, to correct published data we simulated the apparent atmospheric brightness temperature at different airmasses. Then, we applied a non-linear fitting obtaining the apparent zenith opacity and atmospheric temperature.

The results are not significantly different from the 1-2 airmass analytical solution: a minimum apparent opacity has still been found. See Fig. 3, where we simulated an actual observation where skydips were made with measurements at seven different airmasses from 1 to 4 and different opacity values ranging from 0.1 to 4.

We obtained a minimum apparent opacity of 0.6, corresponding to an actual opacity of about 0.2, in good agreement with the minimum value reported at 857 GHz in sites like the South Pole (Radford 2002). The correction to be introduced at high opacity values has also been determined to be about 0.3.

Note that as the response of the instrument is not very steep around $y_{min}$, a small variation in the apparent opacity can correspond to a large variation in the actual value. For this reason, statistics on the atmospheric noise obtained in these conditions should be treated with caution.

Moreover, the value of the reduced $\chi^2$ value rises to very high values at low opacity (for example, 24.5 for $\tau_0 = 0.1$ in the simulation described above), indicating the poor capability of the model (1.1) to explain the observed data and noise distribution.

## CONCLUSIONS

A non-linear error, which becomes very serious at low opacity, is introduced by data reduction algorithms that do not properly take into account the radome transparency.

The consequences are as follow:
- At high opacity values, even if the original data are no longer available and the approximate model (1.1) has been used in data analysis, it is still possible to correct the data by making use of the correcting algorithm (2.10) and (2.11).
- The exact model should be used whenever the apparent atmospheric opacity is equal to or lower than the one given by Fig. 2 at the particular window transparency, in order to remove a possible ambiguity. For example, at 857 GHz ($T_{atm}$=220 K, $T_{win}$= 270 K, transmission=0.81) the exact model (2.5) should be used whenever the apparent opacity at a site falls below ~1. In fact, in this case the approximate model (1.1) fails to produce correct data and statistics. Corrected results at three relevant sites (Radford 2002) are shown in Table 1. Similar corrections apply to other submillimeter windows.
- While the extent of the radome effect depends strictly on the transmission properties of a specific radome material, it is not very sensitive to the radome temperature, as the emitting term is negligible with respect to the absorbing one in eq. (2.5).
- This "radome effect" can explain the "hard lower limit" to the apparent sky opacity observed on some skydip data series obtained at South Pole and at other sites with particularly good atmospheric conditions (Radford 2002; Radford & Holdaway 1998). Significantly, the minimum **detectable** opacity of 0.6 derived from Fig. 2 corresponds to the apparent minimum **detected** opacity for three sites (Chajnantor, South Pole, Mauna Kea).

- The statistics at sites where the opacity is very low (i.e. South Pole in winter) are affected more than those at other sites. For this reason, South Pole could be relatively better than previously reported in terms of average opacity at submillimeter wavelengths.
- Noise statistics would be affected by the window transmission, but in the opposite way: a change in the real opacity does not induce significant variations in the apparent one, so that atmospheric opacity variations will be underestimated. However, this effect is partially mitigated, as seen in statistics on actual data obtained at Dome C (Calisse, et al. 2002), by the fact that the correct model (2.5) fits the real data better, reducing the uncertainty and, in turn, data dispersion.
- Several relations between opacity at submillimeter wavelengths and PWV, obtained by correlating the values of PWV obtained by onsite radiosonde measurements and opacity apparent by radiometers (Chamberlin 2001) have been found. In general, a consistent residual opacity at very low PWV values has been observed and ascribed to a dry-air component. However atmospheric models (Pardo, et al. 2001) predict a much lower level for the dry-air opacity than that measured, particularly at 857 GHz. We believe that this "dry" component has been greatly overestimated in some cases by not correctly taking into account the window properties.

## AKNOWLEDGMENTS

I thank John Storey, Michael Ashley, Michael Burton and Jon Lawrence for useful discussions, tips and comments. In particular, I have to thank John Storey for his endless patience correcting my written *inglese*. I also wish to thank Simon Radford and Jeff Peterson for sharing data from the AST/RO tipper and for some papers in preparation, and Jeff Peterson and Richard Chamberlin for interesting and useful discussions.

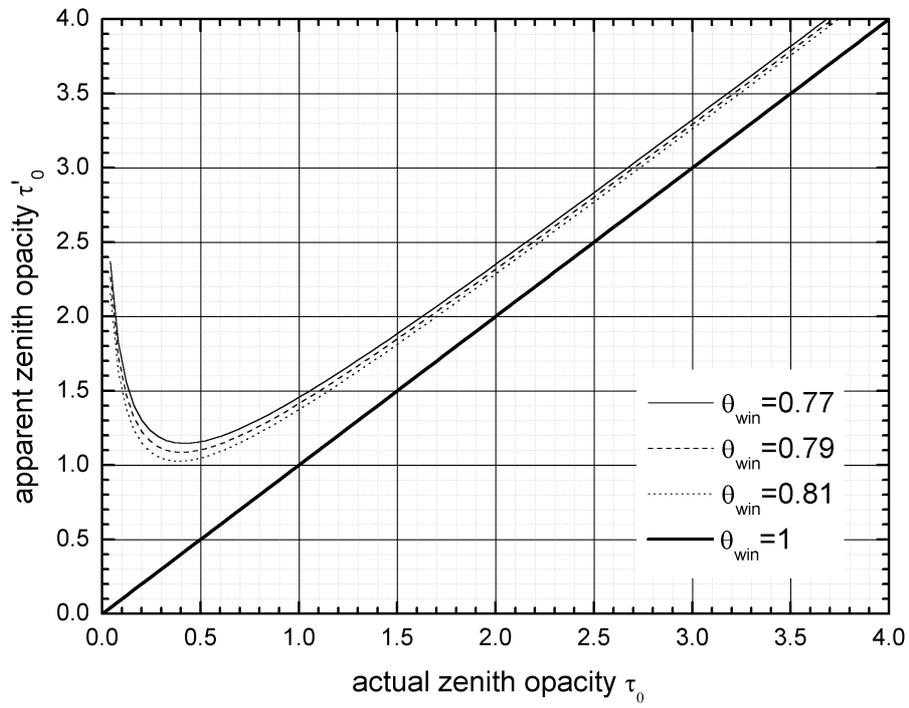

Fig. 1 – Apparent atmospheric opacity as a function of the actual opacity for a range of different window transparencies, when using the analytical solution available for 1 and 2 airmass measurements. In this simulation $T_{atm} = 220K$ and $T_{win} = 270K$.

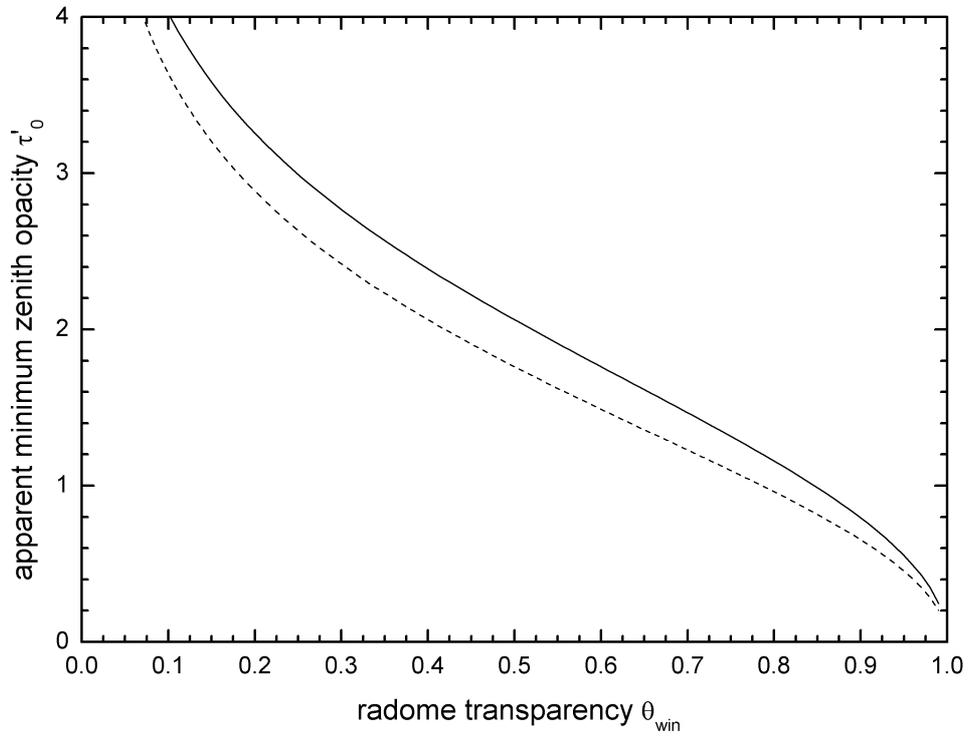

Fig. 2 – Minimum detectable opacity, obtained by the analytical solution of measurements at 1 and 2 airmasses, as a function of the radome transparency. The continuous line represents the case of $T_{win}/T_{atm}=1$, the dashed line represents the case of $T_{win}/T_{atm}=1.5$.

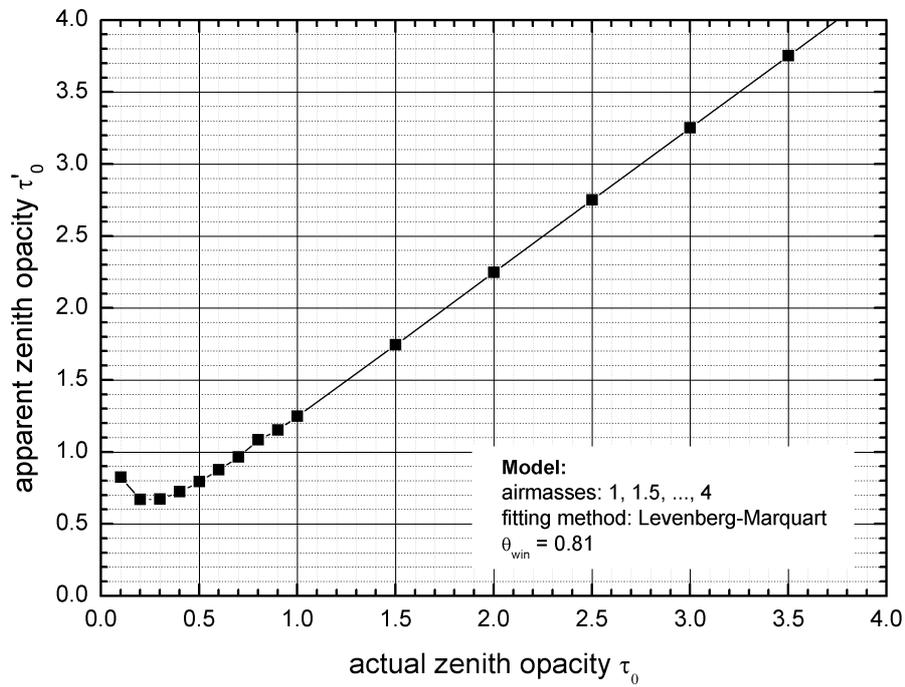

Fig. 3 – Apparent opacity as a function of actual opacity for the case where a non-linear fitting of data to the uniform slab atmospheric model has been used, but the window transparency ignored. In this simulation is $T_{atm} = 220K$ and $T_{win} = 270K$, as in the case of 1-2 airmass (see Fig. 1).

| | 25% | | 50% | | 75% | |
|---|---|---|---|---|---|---|
| | $\tau'_0$ | $\tau_0$ | $\tau'_0$ | $\tau_0$ | $\tau'_0$ | $\tau_0$ |
| Chajnantor | 1.21 | 0.95 | 1.68 | 1.4 | 2.52 | 2.3 |
| Mauna Kea | 1.46 | 1.2 | 2.15 | 1.9 | 3.14 | 2.9 |
| South Pole | 1.23 | 1.0 | 1.52 | 1.3 | 1.94 | 1.7 |

BROADBAND ZENITH OPACITY AT 350 μm

Table 1 – Recalibration of the cumulative percentile distribution of broadband opacity at three relevant sites for submillimeter astronomy, by applying the correction algorithm to results found in the literature (Radford 2002). $\tau'_0$ is the published zenith opacity and $\tau_0$ the actual zenith opacity. The parameters used calculating the correction are shown in Fig. 3.